\documentclass[aps,prd,showpacs,amssymb,floatfix,twocolumn]{revtex4}
\usepackage{graphicx}
\usepackage[sort&compress]{natbib}
\begin{document}
\newcommand{\eg}{{\it e.g.}}
\newcommand{\etal}{{\it et. al.}}
\newcommand{\ie}{{\it i.e.}}
\newcommand{\be}{\begin{equation}}
\newcommand{\dd}{\displaystyle}
\newcommand{\ee}{\end{equation}}
\newcommand{\bea}{\begin{eqnarray}}
\newcommand{\eea}{\end{eqnarray}}
\newcommand{\bef}{\begin{figure}}
\newcommand{\eef}{\end{figure}}
\newcommand{\bce}{\begin{center}}
\newcommand{\ece}{\end{center}}
\def\lsim{\mathrel{\rlap{\lower4pt\hbox{\hskip1pt$\sim$}}
    \raise1pt\hbox{$<$}}}         
\def\gsim{\mathrel{\rlap{\lower4pt\hbox{\hskip1pt$\sim$}}
    \raise1pt\hbox{$>$}}}         

\title{Quark matter in Neutron Stars within the Field Correlator Method}
\author{S. Plumari$^{a,b}$, G. F. Burgio$^c$, V. Greco$^{a,b}$, and  D. Zappal\`a$^c$}
\affiliation{$^a$ Dipartimento di Fisica e Astronomia, Universit\`a di Catania, Via Santa Sofia 64,
I-95123 Catania, Italia}

\affiliation{$^b$ INFN - Laboratori Nazionali del Sud, Via Santa Sofia 62,
I-95125 Catania, Italia}

\affiliation{$^c$ INFN Sezione di Catania, Via Santa Sofia 64,
I-95123 Catania, Italia}

\date{\today}
\begin{abstract}
We discuss the appearance of quark matter in neutron star cores, focussing on the possibility that the recent 
observation of a very heavy neutron star could constrain free parameters of quark matter models.
For that, we use the equation of state derived with the Field Correlator Method, extended to the zero temperature limit, 
whereas for the hadronic phase we use the equation of state obtained within 
both the non-relativistic and the relativistic  Brueckner-Hartree-Fock many-body theory. 
We find a strong dependence of the maximum mass both on the value of the $q\bar q$ interaction $V_1$, 
and on the gluon condensate $G_2$, for which we introduce a dependence on the
baryon chemical potential $\mu_B$. We find that the maximum masses are 
consistent with the observational limit for not too small values of $V_1$.
\end{abstract}

\maketitle

------------------------------------------------------------

\section{Introduction}

The appearance of quark matter in the interior of massive neutron stars (NS) is one of the mostly
debated issues in the physics of these compact objects. Many equations of state (EoS) have been used to describe the interior of NS. If we consider only purely nucleonic degrees of freedom and the EoS is derived within  microscopic approaches \cite{gabri}, it turns out that for the heaviest NS, close to the maximum mass (about two solar masses), the central particle density reaches values larger than $1/fm^3$. In this density range the nucleon cores (dimension $\approx$ 0.5 fm) start to touch each other, and it is hard to imagine that only nucleonic degrees of freedom can play a role. On the contrary, it can be expected that even before reaching these density values, the nucleons start to lose their identity, and quark degrees of freedom are excited at a macroscopic level.

Unfortunately it is not straightforward to predict the relevance of quark degrees of freedom in the interior of NS for the various physical observables, like cooling evolution, glitch characteristics, neutrino emissivity, and so on. The value of the maximum mass of NS is probably one of the physical quantities that is most sensitive to the presence of quark matter in NS. If the quark matter EoS is quite soft, the quark component is expected to appear in NS and to affect appreciably the maximum mass value.
The recent observation of a large NS mass in PSR J0348+0432 with mass $\rm M=2.01 \pm 0.04 M_\odot$ \cite{maxpuls} implies that the EoS of NS matter is stiff enough to keep the maximum mass at these large values. Purely nucleonic EoS are able to accommodate such large masses\cite{gabri}. Since the presence of non-nucleonic degrees of freedom, like hyperons and quarks, tends usually to soften considerably the EoS with respect to
purely nucleonic matter, thus lowering the mass value, their appearance would in this case be incompatible with observations. The large value of the mass could then be explained only if both hyperonic and quark matter EoS are much stiffer than expected. 
Unfortunately, while the microscopic theory of the nucleonic EoS has reached a high degree of sophistication, the quark matter EoS is poorly known at zero temperature and at the high baryonic density appropriate for NS. One has, therefore, to rely on models of quark matter, which contain a high degree of uncertainty. The best one can do is to compare the predictions of different models and to estimate the uncertainty of the results for the NS matter as well as for the NS structure and mass. In this paper we will use two definite nucleonic EoS, which have been developed on the basis of 
the Brueckner-Hartree-Fock many-body theory for nuclear matter, and the Field Correlator Model (FCM) for the quark EoS \cite{phrep},
which in principle is able to cover the full temperature-chemical potential plane.
The FCM EoS contains {\it ab initio} the property of
confinement, which is expected to play a role as far as the stability of a neutron star is concerned
\cite{noi07}, at variance with other models like, e.g., the Nambu--Jona-Lasinio model.
In a previous  paper \cite{noi08} we analyzed the EoS of the quark matter in the FCM
(for a review see \cite{phrep}), and found that the model could be tested against NS observations,
and these could seriously constrain the parameters used in the model.
It was shown that this approach admits stable NS with gravitational masses slightly larger than 1.44 $M_\odot$, thus providing numerical indications on some relevant
physical quantities, such as the gluon condensate. 
In the present paper, we elaborate further on this idea, and explore the dependence of the model
on the $q\bar q$ potential $V_1$ and, moreover, the dependence of the gluon condensate
$G_2$ on the baryon chemical potential $\mu_B$. The observation of a very large neutron star mass \cite{maxpuls} can be used to put constraints on these two parameters.
 
This paper is organized as follows : in the next Section the FCM at finite temperature and density is briefly recalled, with an extensive discussion of the model parameters, while Sec.III  contains some details of the EoS for the hadronic phase. 
In Sec.IV the hadron-quark phase transition is illustrated and the  results of our analysis are presented in Sec.V.  Finally, Sec.VI is devoted to the conclusions. 

\section{Quark Matter: EoS in the Field Correlator Method}

The approach based on the FCM  provides a natural treatment
of the dynamics of confinement in terms of the
Color Electric   ($D^E$ and $D_1^E$)  and Color Magnetic ($D^H$ and $D_1^H$) 
Gaussian correlators, being the former one directly related to confinement,
so that its vanishing above the critical temperature implies deconfinement \cite{phrep}.
The extension of the FCM to finite temperature $T$ and chemical potential $\mu_q=0$ 
gives  analytical results in reasonable agreement with lattice data 
thus allowing  to describe correctly the deconfinement phase transition
\cite{sim1,sim4,sim22,sim5,sim55,sim6}.  In this work, we are interested in the physics of 
neutron stars, and therefore the extension of the FCM to finite values of the chemical potential \cite{sim4,sim22} 
allows to obtain the Equation of State of the quark-gluon matter in the 
range of baryon density typical of the neutron star interiors. 

Within the FCM, the quark pressure for a single flavour is simply given by \cite{sim4,sim22,sim6}
\be\label{pquark}
P_q/T^4 = \frac{1}{\pi^2}[\phi_\nu (\frac{\mu_q - V_1/2}{T}) +
\phi_\nu (-\frac{\mu_q + V_1/2} {T})]
\ee
where 
\be
\phi_\nu (a) = \int_0^\infty du \frac{u^4}{\sqrt{u^2+\nu^2}} \frac{1}{(\exp{[ \sqrt{u^2 +
\nu^2} - a]} + 1)}.
\ee
being $\nu=m_q/T$,  and $V_1$  the large distance static $q \bar q$ potential:
\be
\label{v1}
V_1 = \int_0^{1/T} d\tau(1-\tau T) \int_0^\infty d\chi \chi D_1^E(\sqrt{\chi^2 + \tau^2})
\ee
The potential  $V_1$ in Eq.(\ref{v1}) is assumed to be independent on the
chemical potential, and this is partially supported by lattice simulations
at very small chemical potential \cite{sim22,latmuf}. We elaborate more on this point in the 
following subsection.

The EoS is completely specified once the gluon contribution is added to the quark pressure,
i.e.
\be\label{pglue}
P_g/T^4 = \frac{8}{3 \pi^2} \int_0^\infty  d\chi \chi^3
\frac{1}{\exp{(\chi + \frac{9 V_1}{8T} )} - 1}
\ee
and therefore
\be
\label{pqgp1}
P_{qg} =P_g+\sum_{j=u,d,s} P^j_{q} + \Delta \epsilon_{vac}
\ee
where $P_q$ and $P^j_{g}$ are respectively given in Eq. (\ref{pquark}) and
(\ref{pglue}), and
\be
\label{pqgp2}
\Delta \epsilon_{vac}
\approx - \frac{(11-\frac{2}{3}N_f)}{32} \frac{G_2}{2}
\ee
corresponds to the difference of the vacuum energy density in the two phases,
being $N_f$ the flavour number. $G_2$ is the gluon condensate whose numerical value, 
determined by the QCD sum rules, is known with large uncertainty \cite{gluecond}
\be
G_2=0.012\pm 0.006~ \rm{GeV^4}
\ee

Therefore the EoS in  Eq.(\ref{pqgp1}) essentially depends on two parameters,
namely the quark-antiquark potential $V_1$ and the gluon condensate $G_2$.
In addition, at finite temperature 
and vanishing baryon density, a comparison with the available lattice calculations of the Wuppertal-Budapest, \cite{wup_buda,wup_buda_2},
and hotQCD collaborations, \cite{hotqcd,hotqcd_2,hotqcd_new}, 
provides clear indications  about the specific values of these parameters, and in particular their values at the critical temperature $T_c$.
These estimates are related to the corresponding values of the parameters at $T=\mu_B=0$
which, in turn,  can be used as an input to study  the EoS at $T=0$ and finite $\mu_B$.  

\subsection{The $\rm V_1$ and $\rm G_2$ parameters}
\label{subsec}

In ref.\cite{sim4} the EoS at zero baryon density has been derived,  by explicitly assuming a temperature 
dependence of the gluon condensate $G_2$ as found 
in  lattice simulations \cite{elia1,elia2}, namely an almost constant $G_2(T)$ for $0<T<T_c$, with a sudden drop 
around $T_c$ to one half of its value, followed by the constant behavior  $G_2 (T) = G_2(T=0) /2$, for $T>T_c$.  In addition,  an indication on the value of $V_1(T_c)$ 
has been extracted in \cite{bombaci}, starting from the expression  of the critical temperature obtained  in \cite{sim4,sim22}
\be
\label{tcritical}
T_c=\frac{a_0 G_2^{1/4}}{2} \left (1 +\sqrt{ 1 + \frac {V_1(T_c)} {2a_0 G_2^{1/4}  }}  \right )\; ,
\ee
where $a_0=(3 \pi^2 / 768)^{1/4}$.
In fact, once the values of $G_2$ and $T_c$ are fixed, 
one  immediately gets $V_1(T_c)$  from Eq. (\ref{tcritical}), and in ref. \cite{bombaci} it has been
shown that, for $G_2(T=0) =0.012 \, \rm{GeV}^4$, the critical temperatures found in  \cite{wup_buda,hotqcd_2}, respectively $\rm T_c=147 \pm 5 ~MeV$ and $\rm T_c=154 ~\pm~ 9 ~MeV$,
correspond to rather small values of  $V_1$  ($V_1(T_c) \lesssim 0.15$ GeV),  
while the optimum value indicated in \cite{sim22}, $V_1(T_c) =0.5$ GeV,  reproduces 
those temperatures for small values of $G_2$, i.e. $G_2 \simeq 0.004 \, \rm{GeV^4} $.

\begin{figure*}[t] 
\centering
\includegraphics[width=8.5cm,angle=0]{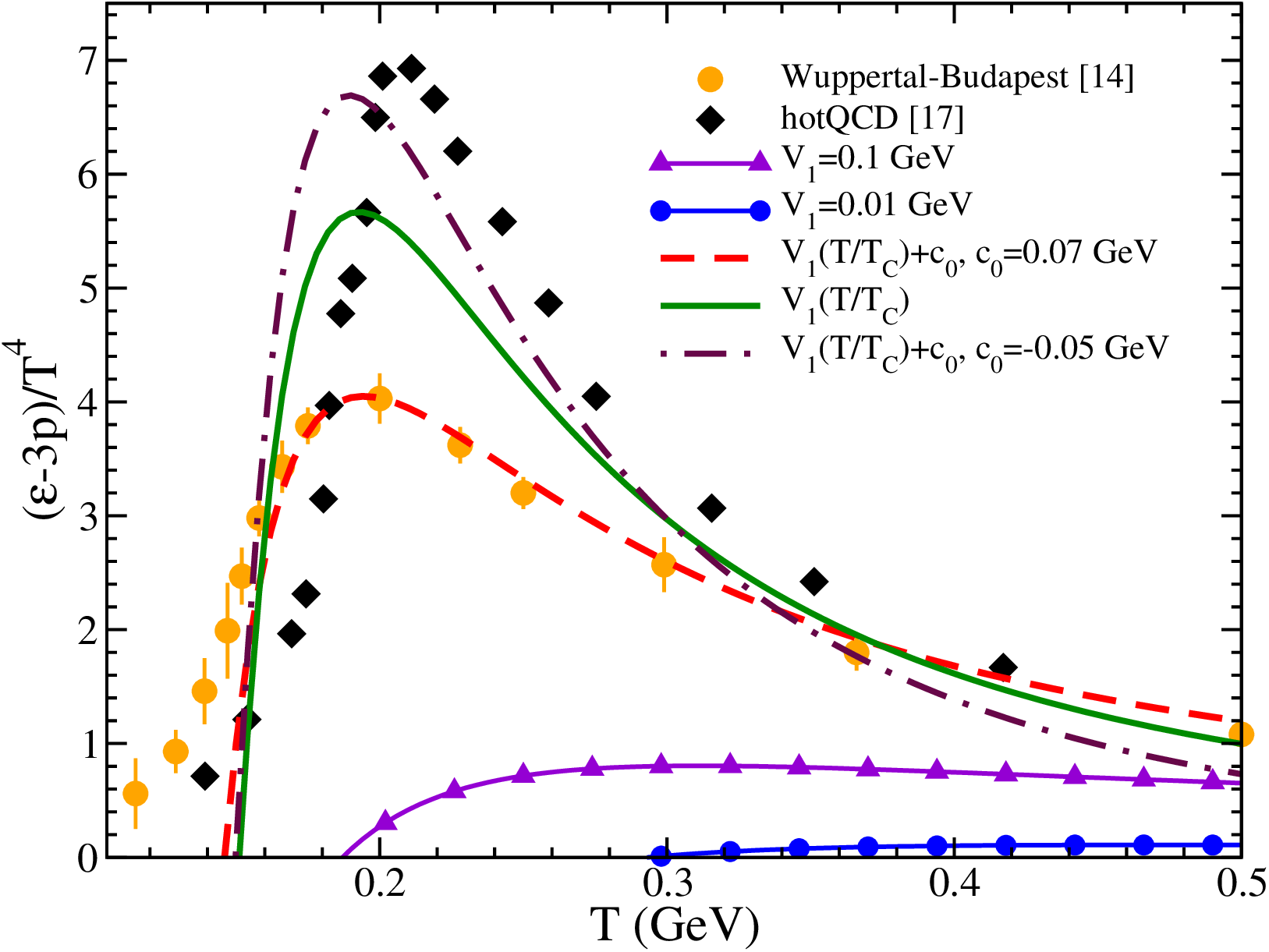}
\caption{ The interaction measure  $\rm (\epsilon-3p)/T^4$ as a function of the temperature as 
obtained in the FCM for three values of $c_0$ in Eq. (\ref{v1t}) : $c_0= 0.07,\, 0,\, -0.05$ GeV (respectively: dashed (red), solid
(green)  and dot-dashed (brown) curve), compared with the lattice data  of ref. \cite{wup_buda_2} (orange circles) and ref. \cite{hotqcd} (black , diamonds).}
\label{f:finitet}
\end{figure*} 

However,  one should recall that Eq.(\ref{tcritical}) is not extremely accurate, being  obtained 
by neglecting the hadron pressure at the transition, which in \cite{sim22} is estimated as a $10\%$ uncertainty.
Hence a check of the EoS focused on the critical point $T=T_c$ only, 
could be too restrictive, as the numerical data on lattice cover a large temperature range above $T_c$. 
For that,  we compare in Fig.\ref{f:finitet} the predictions 
of the FCM with the available  lattice data  around and above the critical temperature. 
In Fig.\ref{f:finitet} we concentrate on the interaction measure $(\epsilon-3p)/T^4$, which is particularly significant because it depends both on the energy density and on the pressure of the system and shows, around the critical temperature, large deviations from zero, i.e. the value of the interaction measure of a free gas of massless particles.
The predictions of the FCM are checked against the lattice data for different parametrizations of $V_1(T)$, and also for constant $\rm V_1=0.01 \, GeV$ and $\rm V_1=0.1\, GeV$. From Fig. \ref{f:finitet} it is evident that these two constant values are too small, and  higher values of $V_1$ must be considered. 
As suggested in ref.\cite{sim22}, we took  
\be\label{v1t}
V_1(T)=c_0 + V_1 \left(\frac{T}{T_c} \right) =c_0 + 0.175 \,  \left ( 1.35 \frac{T}{T_c} -1 \right )^{-1} \;  \rm{ GeV} 
\ee
with respectively $\rm c_0=0.07,\, 0,\, -0.05$ GeV  (corresponding to  $\rm V_1(T_c)= 0.57,\, 0.5,\, 0.45$ GeV ).
The results, displayed by the dashed (red), solid (green)  and dot-dashed (brown) curves,
show a much better agreement with the data, and in particular the dashed (red) curve with 
$\rm V_1(T_c)=0.57$ GeV gives 
a good fit to the data of \cite{wup_buda_2}, represented by full circles (orange), whereas $\rm V_1(T_c)=0.45$ is preferable for the data in \cite{hotqcd}
from the hotQCD collaboration, and shown as full diamonds (black).
However, we warn that, more recently, also the hotQCD collaboration is converging toward a smaller peak for $(\epsilon - 3P)/T^4 $ close to the Wuppertal-Budapest one as presented in \cite{hotqcd_new}.

This check shows that the analysis of the  whole set of lattice data  points toward  a value of $V_1(T_c)$ around   $0.5 - 0.6$ GeV while, 
as noticed above, the simple determination of  the critical temperature $T_c$ with reasonable values of $G_2$  would 
suggest smaller  $V_1(T_c)$. The value of the potential at $T=0$, which is essential, for the study of NS structure,  
has been computed in \cite{bombaci} as a function of $V_1(T_c)$, by making use of Eq. (\ref{v1}), under the assumption of a temperature 
independent $D_1^E$ in the region $0<T<T_C$ \cite{elia2}, obtaining $V_1(T=0) \simeq 0.8 \div 0.9$ GeV in correspondence of  $V_1(T_c)=0.5$ GeV.

It is important to notice that there is no direct relation of these values with the potential at finite 
$\mu_B$. One would expect that an increasing
baryon density could produce a screening effect that reduces the intensity of the  quark-antiquark potential, and at large density the 
quark-quark 
interaction should become more and more relevant. In our analysis we choose to 
keep  $V_1$ as a free parameter, and check what kind of indications on 
$V_1$ can be extracted from the determination of the maximum mass of neutron stars.

Let us now turn to the other parameter of the FCM model, namely the gluon condensate $G_2$. As mentioned above,
$G_2(T)$ at zero baryon density has been computed on lattice  \cite{elia1,elia2}  but, due to technical difficulties, 
analogous calculations in full QCD at large  $\mu_B$ are  precluded. 
Therefore we have to resort to different approaches to get some indications on 
the gluon condensate at $\mu_B\neq 0$. In particular, the QCD sum rules technique has been used to study
some hadronic properties within a nuclear matter environment at $T=0$ \cite{cohen}, and it has been found that the gluon 
condensate decreases linearly with the baryon density $\rho_B$ ($m_N$ indicates the neutron mass) 
\be\label{g2ro}
G_2(\rho_B) - G_2(\rho_B=0) = - m_N \rho_B + O(\rho_B^2) \, .
\ee
Further analysis \cite{druc,balcasza} show that the corrections to Eq. (\ref{g2ro}), 
even  when including  nonlinear effects, 
are substantially small 
and can be neglected for  our purposes.  According to this decreasing trend,  the  gluon condensate vanishes at some value of the baryon density
and, as noticed in \cite{balcasza} one expects that a transition to the deconfined state should  occur before reaching this point.

Once the behaviour of the condensate below the transition is given in Eq. (\ref{g2ro}), we still 
need to establish  $G_2(\mu_B) $ at higher values of the baryon chemical potential to proceed in our analysis. 
Rather than following the  simplest choice
of taking an effective $\mu_B$-independent $G_2$,  which was adopted in \cite{noi08,bombaci}, we prefer to retain 
Eq. (\ref{g2ro}) at lower densities and, at the same time, to follow the indications  proposed in \cite{zit1,zit2} at higher density.
In fact, in \cite{zit1,zit2} it is suggested that $G_2(\mu_B) $ in full three-color ($N_c=3$) QCD, has the same qualitative behavior 
of the corresponding variable in two-color ($N_c=2$) QCD. In this case many technical problems that affect the theory with $N_c=3$, are absent
and, in particular, the modification of the gluon condesate at finite chemical potential, namely the difference 
$f_{CS}(\mu)=G_2(\mu) - G_2(0)$, is computed from the energy momentum tensor of an effective chiral lagrangian with the following result:
\be\label{g2qu}
f_{CS}(\mu)=
4f_\pi^2 ( \mu^2 -M^2   ) \left (  1 - \frac{M^2}{\mu^2}    \right )
\ee
where $M$ is identified with the pion mass.
Eq. (\ref{g2qu}) shows an initial decrease which, after reaching a minimum, is followed by a continuous growth.
This trend is understood with the 
appearance of a weakly interacting gas of diquarks, 
whose pressure is negligible if compared to its energy density, which mostly comes
from diquark rest mass. Accordingly, the gluon condensate, that is related to minus the trace energy momentum tensor,
decreases with $\mu_B$.
Only at sufficiently large chemical potential 
the contribution of the diquarks on the pressure becomes approximately 
equal to  energy density and the  growth of the gluon condensate is observed.
This result  is also consistent with lattice calculations   \cite{sands,alles}  which can be carried out for $N_c=2$. 
Finally in \cite{zit1,zit2} it is claimed that the color superconducting (CS) phase in the $N_c=3$ theory
should  qualitatively reproduce the picture described above, and Eq. (\ref{g2qu}) does still hold, 
provided that in this case 
one takes $M \sim 2 \Lambda_{QCD}$ and $\mu$
is identified with the quark chemical potential : $\mu=(1/3) \mu_B$.

\begin{figure}[t] 
\centering
\includegraphics[width=8.5cm,angle=0]{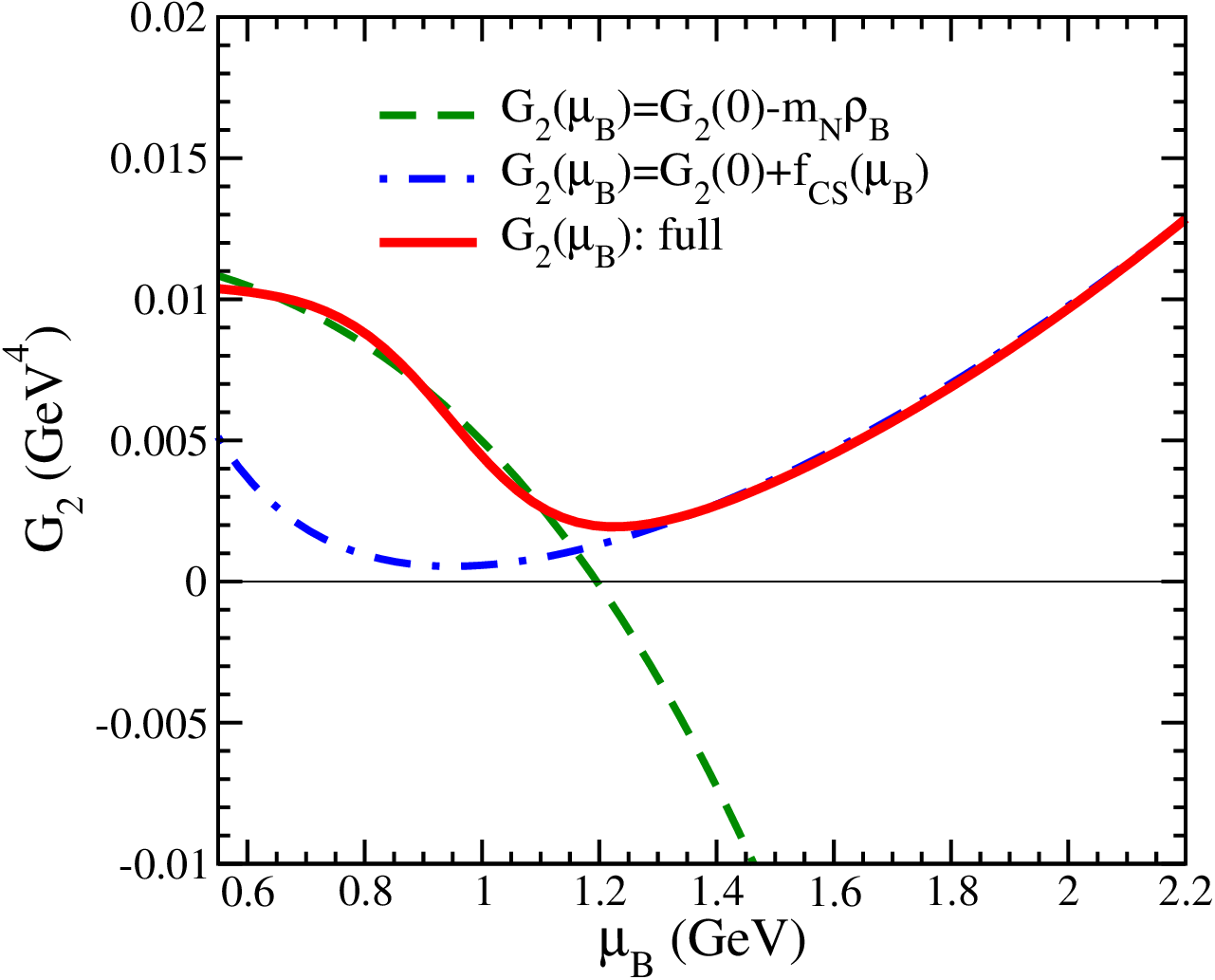}
\caption{$G_2(\mu_B)$ as obtained from Eq. (\ref{g2ro})  (green dashed) 
and Eq. (\ref{g2qu}) (blue dot-dashed) with $G_2(\mu_B=0)=0.012$ GeV$^4$. The solid (red) line is the effective approximation used in our analysis.  (see text). }
\label{f:g2mu}
\end{figure} 

Therefore we have put together the two curves of the gluon condensate at low and at high values
 of $\mu_B$, given respectively  in Eqs. (\ref{g2ro}) and (\ref{g2qu}), 
and selected $G_2(\mu_B)$ as the solid (red)  line displayed in Fig. \ref{f:g2mu}. 
More precisely, in Fig. \ref{f:g2mu} the dashed (green) and the dot-dashed (blue) lines  respectively correspond 
to  Eq. (\ref{g2ro}) (where $G_2(\rho_B)$ is reparametrized in terms of $G_2(\mu_B)$) 
and Eq.  (\ref{g2qu}). Then, by crudely assuming that the transition point lies close to 
the intersection point of these two curves, we parametrized $G_2(\mu_B)$ through an effective analytic expression, the solid (red) line,
which  approximates  Eq. (\ref{g2ro})  at low $\mu_B$ and Eq.  (\ref{g2qu}) at higher $\mu_B$.
This analytic form avoids a discontinuity in the derivative of $G_2(\mu_B)$ at the intersection of the two curves 
that could produce unphysical features  when computing  the pressure or the energy density of the system.
In Fig. \ref{f:g2mu} the value of the condensate at  zero chemical potential  is  taken $G_2(\mu_B=0)=0.012$ GeV$^4$.

\section{Hadronic Phase: EoS in the Brueckner-Bethe-Goldstone theory}

In this section we remind briefly the BHF method for the nuclear matter EoS. 
This theoretical scheme is based on the Brueckner-Bethe-Goldstone (BBG) many-body theory, 
which is the linked cluster expansion of the energy per nucleon of nuclear matter 
(see Ref.\cite{book}, chapter 1 and references therein). In this many-body approach one systematically replaces the bare nucleon-nucleon (NN) interaction V by the Brueckner reaction matrix G, which 
is the solution of the Bethe-Goldstone equation

\begin{eqnarray}
G(\rho;\omega) =  V  + V \sum_{k_a k_b} {{|k_a k_b\rangle  Q  \langle k_a k_b|}
  \over {\omega - e(k_a) - e(k_b) }} G(\rho;\omega),
\end{eqnarray}
\noindent
where $\rho$ is the nucleon number density, $\omega$  is the  starting energy, and
$|k_a k_b\rangle Q \langle k_a k_b|$  is  the Pauli operator.
$e(k) = e(k;\rho) = {{\hbar^2}\over {2m}}k^2 + U(k;\rho)$
is the single particle energy,  and $U$ is the single-particle potential,

\begin{eqnarray}
U(k;\rho) &= &\sum _{k'\leq k_F} \langle k k'|G(\rho; e(k)+e(k'))|k k'\rangle_a
\end{eqnarray}
The subscript ``{\it a}'' indicates antisymmetrization of the
matrix element. In the BHF approximation the energy per nucleon is
 \begin{eqnarray}
&&{E \over{A}}(\rho)  =
          {{3}\over{5}}{{\hbar^2~k_F^2}\over {2m}} + D_{\rm 2}\, , \\
&&D_{\rm 2} = {{1}\over{2A}}
\sum_{k,k'\leq k_F} \langle k k'|G(\rho; e(k)+e(k'))|k k'\rangle_a
\end{eqnarray}
\noindent

The nuclear EoS can be calculated with good accuracy in the Brueckner two hole-line 
approximation with the continuous choice for the single-particle potential, since the results in 
this scheme are quite close to the calculations which include also the three hole-line 
contribution. However, as it is well known, the non-relativistic calculations, based on 
purely two- body interactions, fail to reproduce the correct saturation point of symmetric 
nuclear matter and one needs to introduce three- body forces (TBFs). In our approach the 
TBF's are reduced to a density dependent two-body force by averaging over the position of the third particle \cite{bbb}

In this work we choose the Argonne $v_{18}$ nucleon-nucleon potential \cite{v18}, supplemented  by the so-called Urbana model \cite{uix} as three-body force.
This allows to reproduce correctly the nuclear matter saturation point
$\rho_0 \approx 0.17~\mathrm{fm}^{-3}$, $E/A \approx -16$ MeV, and gives
values of incompressibility and symmetry energy at saturation compatible
with those extracted from phenomenology \cite{myers}. For completeness we will show results 
obtained with the relativistic counterpart, i.e. the Dirac-Brueckner-Hartree-Fock scheme \cite{fuchs}
where the Bonn A potential is used as NN interaction.
In the low density region ($\rho \rm < 0.3~ fm^{-3}$), both BHF+TBF binding energies and DBHF calculations are very similar, whereas at higher densities the DBHF is slightly stiffer  \cite{gabri}. 
The discrepancy between the nonrelativistic and relativistic calculation can be easily understood by noticing that the DBHF treatment is equivalent  to introducing in the nonrelativistic BHF 
the three-body force corresponding to the excitation of a nucleon-antinucleon pair, the so-called Z-diagram \cite{Z_diag}, which is repulsive at all densities. On the contrary, in the BHF treatment both attractive and repulsive three-body forces are introduced, and therefore a softer EoS is expected.

We remind that the BBG approach has been extended to the hyperonic sector
in a fully self-consistent way \cite{hypmat,hypns}, by including the
$\Sigma^-$ and $\Lambda$ hyperons, but in this paper we consider stellar matter as
composed by neutrons, protons, and leptons in beta equilibrium \cite{bbb}.
The chemical potentials of each species are the fundamental input for solving
the equations of chemical equilibrium, charge neutrality and baryon number conservation,
yielding the equilibrium fractions of all species. 
Once the composition of the $\beta$-stable, charge neutral stellar matter
is known, one can calculate the equation of state, i.e., the relation between
pressure $P$ and energy density $\epsilon$ as a function of the baryon density
$\rho$. It can be easily obtained from the thermodynamical relation
\begin{eqnarray}
P &=& - \frac{dE}{dV} = P_B + P_l \label{phad1}\\
P_B &=& \rho^2 \frac{d(\epsilon_B/\rho)}{d\rho}, \;\; \;
P_l = \rho^2 \frac{d(\epsilon_l/\rho)}{d \rho}\label{phad2}
\end{eqnarray}
with $E$ the total energy and $V$ the total volume. The total nucleonic energy
density is obtained by adding the energy densities of each species
$\epsilon_i$. As far as
leptons are concerned, at those high densities electrons are a free
ultrarelativistic gas, whereas muons are relativistic. Hence their energy
densities $\epsilon_l$ are well known from textbooks \cite{shapiro}.
The numerical procedure has been often illustrated in papers and textbooks \cite{shapiro}, 
and therefore it will not be repeated here.

\section{The hadron-quark phase transition}
\label{s:hq}

We are now able to compare the pressure of the two phases, namely
the pressure in the hadronic phase given in Eqs.(\ref{phad1})-(\ref{phad2}), and
the quark pressure shown in Eq.(\ref{pqgp1}).
We adopt the simple Maxwell construction, by assuming a first order hadron-quark phase transition \cite{fodor} in beta-stable matter. The more general Gibbs
construction \cite{gle} is still affected by many theoretical
uncertainties \cite{mixed}, and in any case the final mass-radius
relation of massive neutron stars \cite{mit} is slightly affected.

\begin{figure}[t] 
\centering
\includegraphics[width=8.5cm,angle=0]{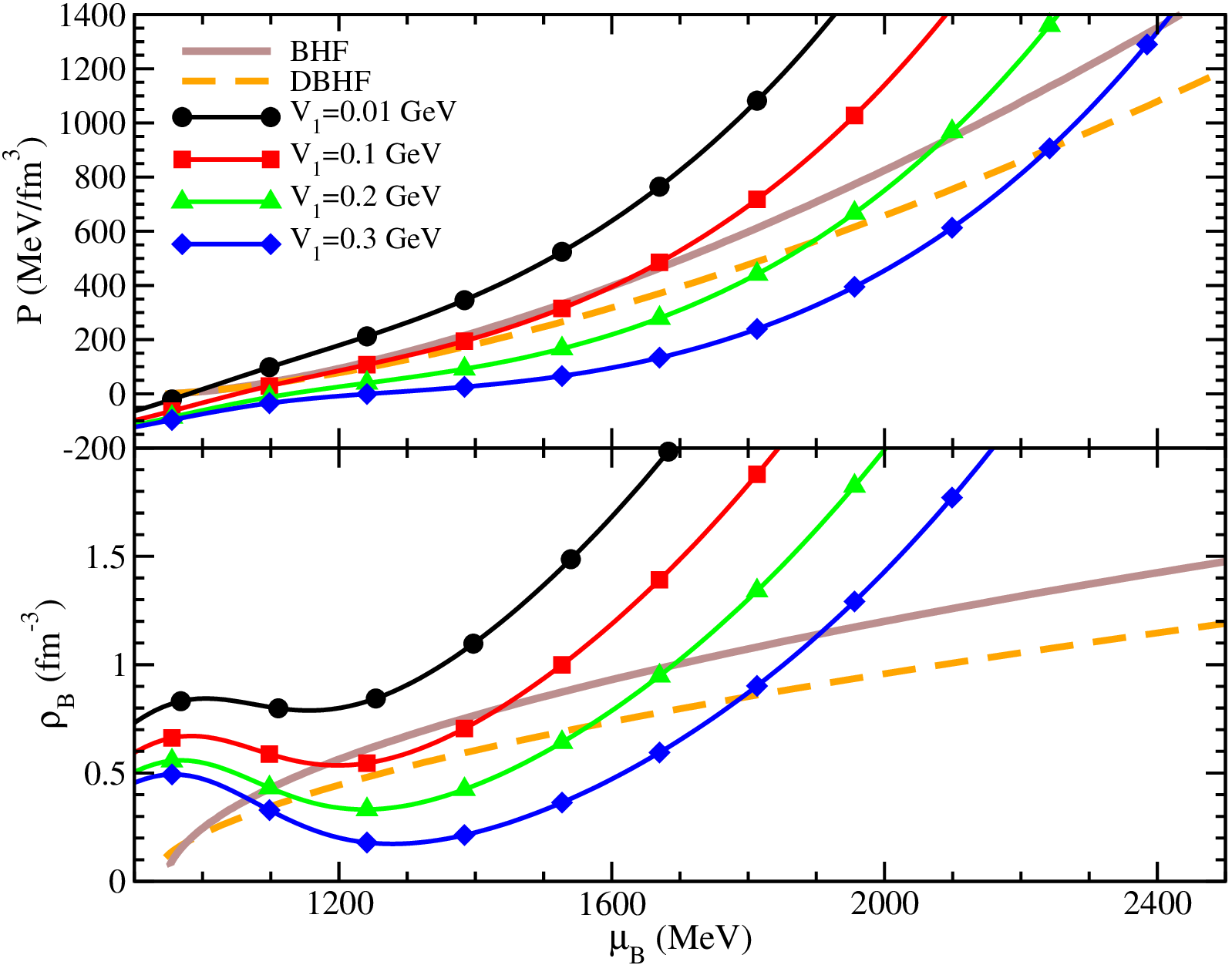}
\caption{The pressure P is displayed vs. the baryon chemical potential in the upper panel,
whereas in the lower panel the baryon density is shown for different values of $V_1$ in the 
FCM model. The EoS's for the hadronic phase are represented by the solid line (brown color) for the BHF, 
and for the DBHF (dashed yellow curve) EoS's. }
\label{f:pmu}
\end{figure} 

We impose thermal, chemical, and mechanical equilibrium between the two phases. This implies
that the phase coexistence is determined by a crossing point in the
pressure vs. chemical potential plot, as shown in Fig.~\ref{f:pmu}.
There we display the pressure
$P$ (upper panel) and the baryon density (lower panel) as function of the baryon chemical potential 
$\mu_B$ for the baryonic and quark matter phases. The hadronic EoS's are plotted as 
solid (brown, BHF), and dashed (orange, DBHF) curves, whereas symbols are the
results for quark matter EoS in the FCM and different choices of $\rm V_1$.
We observe that the crossing points are significantly affected by the
chosen value of the potential $V_1$. Moreover, with increasing $V_1$, the onset of the phase
transition is shifted to larger chemical potentials. Hence, we expect that
the neutron star will possess a thicker hadronic layer with increasing $V_1$.

\begin{figure}[t] 
\centering
\includegraphics[width=8.5cm,angle=0]{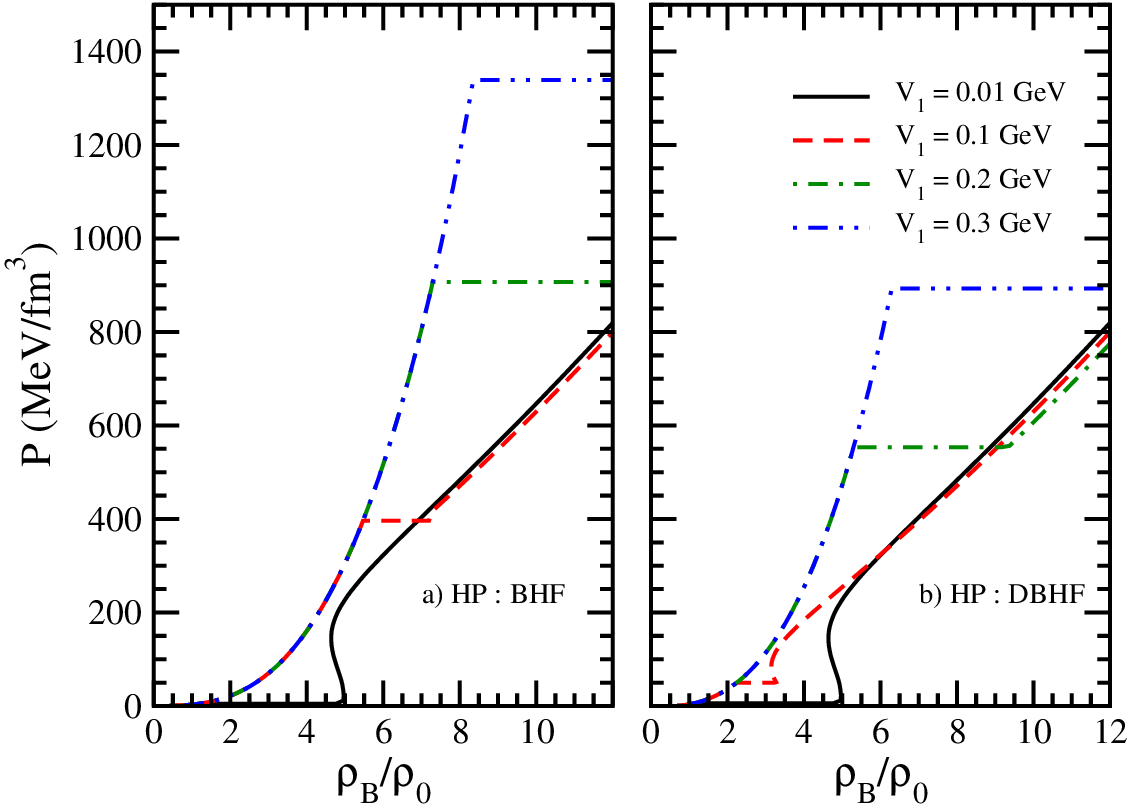}
\caption{The pressure P is displayed vs. the baryon density for different values
of $V_1$ in the FCM model. In the left (right) panel calculations are shown
when the BHF (DBHF) approach is used for the hadronic phase.}
\label{f:EoS}
\end{figure}

In Fig.\ref{f:EoS} we display the total EoS, i.e. the pressure as a function of the
baryon density for the several cases discussed above.
In particular we plot in the left panel the EoS obtained by
using the BHF approach for the hadronic phase, whereas 
in the right panel calculations are shown for the case when
the DBHF EoS is adopted. The several curves represent 
different choices of the $q\bar q$ potential $\rm V_1$. 
The plateaus are
consequence of the Maxwell construction. Below the plateau,
$\beta$-stable and charge neutral stellar matter is in the purely
hadronic phase, whereas for density above the ones characterizing the
plateau, the system is in the pure quark phase.
The main features of the phase transition are displayed in Table \ref{t:phtr}, 
where we report for a fixed hadronic EoS and several values of the 
$q\bar q$ potential $\rm V_1$, the baryonic chemical potential
at the transition $\mu_B^{tr}$, and the corresponding baryon density 
$\rho^{tr}$ in units of the saturation density $\rho_0$, and the gluon condensate $\rm G_2^{tr}$. 
We see that, whenever the transition takes place at $\mu_B \lesssim 1.2~GeV$, i.e. below the minimum shown in Fig.\ref{f:g2mu}, the pressure P 
shows a kink as a function of the baryon density, and this is due to the particular parametric form of the gluon condensate $G_2$ shown in Fig.\ref{f:g2mu}.   The presence of a kink gives unstable neutron stars configurations, as it will be shown in the next Section.

\begin{table}[t]
\setlength{\tabcolsep}{5pt}
\caption{Properties of the hadron-quark phase transition.}
\begin{ruledtabular}
\begin{tabular}{c | c c c c c }
&\multicolumn{1}{c}{$\rm V_1 (GeV)$}
&\multicolumn{1}{c}{$\rm \mu_B^{tr} (GeV)$}
&\multicolumn{1}{c}{$\rm \rho^{tr}/\rho_0$}
&\multicolumn{1}{c}{$\rm G_2^{tr} (GeV^4)$} \\
\hline
      & 0.01  & 0.986 & 1.22 & 0.0048  \\
 BHF & 0.1  & 1.606 & 5.51 & 0.0046  \\
       & 0.2  & 2.09 & 7.33 & 0.0109  \\
      & 0.3  & 2.408 & 8.4 & 0.0165  \\
\hline
        & 0.01 &  0.987 & 1.17 & 0.0047   \\
 DBHF  & 0.1 & 1.14 & 2.26 & 0.0023  \\
   & 0.2   & 1.89 & 5.29 & 0.0079   \\
 	   & 0.3 & 2.24  & 6.31 & 0.0165  \\
\hline
\end{tabular}
\end{ruledtabular}
\label{t:phtr}
\end{table}

\section{Results and discussion}

\begin{figure}[t] 
\centering
\includegraphics[width=7.5cm,angle=0]{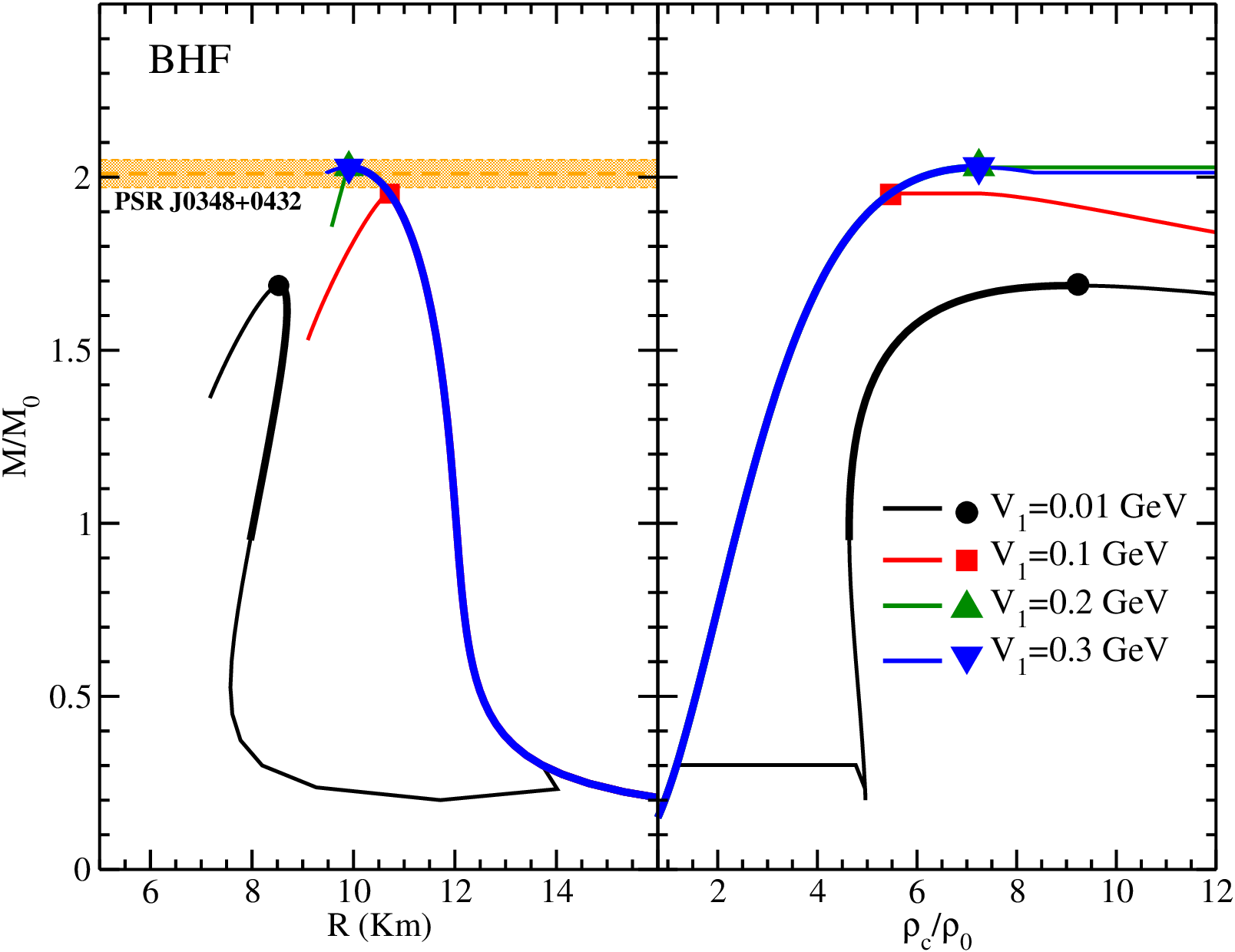}
\caption{The mass-radius (left panel) and the mass-central density relation (right panel) are
 displayed for different values of $V_1$ and the BHF hadronic EoS. The full symbols denote
 the value of the maximum mass. Stable configurations are displayed by thick lines, whereas thin lines indicate unstable configurations. $G_2$ is dependent on $\mu_B$. (see text for details). }
\label{f:mrbhf_g2}
\end{figure}

The EoS is the fundamental input for solving the well-known hydrostatic equilibrium equations of Tolman, Oppenheimer, and Volkoff \cite{shapiro} for the pressure $P$ and the enclosed mass $m$
\bea
 {dP(r)\over dr} &=& -\frac{Gm(r)\epsilon(r)}{r^2}
 \frac{\big[ 1 + {P(r)\over\epsilon(r)} \big]
       \big[ 1 + {4\pi r^3P(r)\over m(r)} \big]}
 {1-{2Gm(r)\over r}} \:,
\label{tov1:eps}
\\
 \frac{dm(r)}{dr} &=& 4\pi r^{2}\epsilon(r) \:,
\label{tov2:eps}
\eea
being $\epsilon$ the total energy density ($G$ is the gravitational constant).
For a chosen central value of the energy density, the numerical integration of
Eqs.~(\ref{tov1:eps}) and (\ref{tov2:eps}) provides the mass-radius relation.
For the description of the neutron star crust, we have joined the equations of state above described with the ones by Negele \& Vautherin \cite{nv} in the medium-density regime,
and the ones by Baym, Pethick, \& Sutherland \cite{bps}
for the outer crust ($\rho<0.001\;\mathrm{fm}^{-3}$), and Feynman, Metropolis, \& Teller \cite{fmt}. 
In Fig.\ref{f:mrbhf_g2} we display the gravitational mass
(in units of solar mass $\rm M_\odot = 2\times 10^{33}g$) as a
function of the radius $\rm R$ (left panel) and the corresponding central baryon density,
normalized with respect to the saturation value (right panel). Stellar configurations have been
obtained using the BHF EoS for the hadronic phase. The orange band represents the recently observed neutron star  PSR J0348+0432 with mass $\rm M = 2.01 \pm 0.04 ~M_\odot$ \cite{maxpuls}.  We have marked the stable configurations by thick lines, whereas full symbols denote the maximum mass. Unstable 
configurations are displayed by thin lines. Among the unstable configurations, we signal those
characterized by increasing mass and decreasing central density, which are related to the appearance 
of the kink in the EoS, as anticipated in Sect.\ref{s:hq}.
In Table \ref{t:mass} we display the values characterizing the maximum
mass, i.e. the central density (in units of $\rho_0$) and the corresponding value of the gluon
condensate $\rm G_2$ whenever the core contains quark matter. Those configurations are stable and the mass values are denoted by an asterisk. We see that the maximum value spans over a range between 1.69 and 2.03  solar masses, depending on the value of the $q\bar q$  potential  $V_1$.  However, the observational data indicate that values of $\rm V_1 $ as small as  $\rm 0.01 ~GeV$ are excluded, and that values of about $\rm 0.1~GeV$ are only marginally compatible with the observational data.  Larger values of $V_1$ produce increasing maximum masses, but the 
stable stars are in the purely hadronic phase. We found that $\rm V_1 \approx 0.095~GeV$ is the largest value which produces stable neutron stars with a quark core. 

\begin{figure}[t] 
\centering
\includegraphics[width=7.5cm,angle=0]{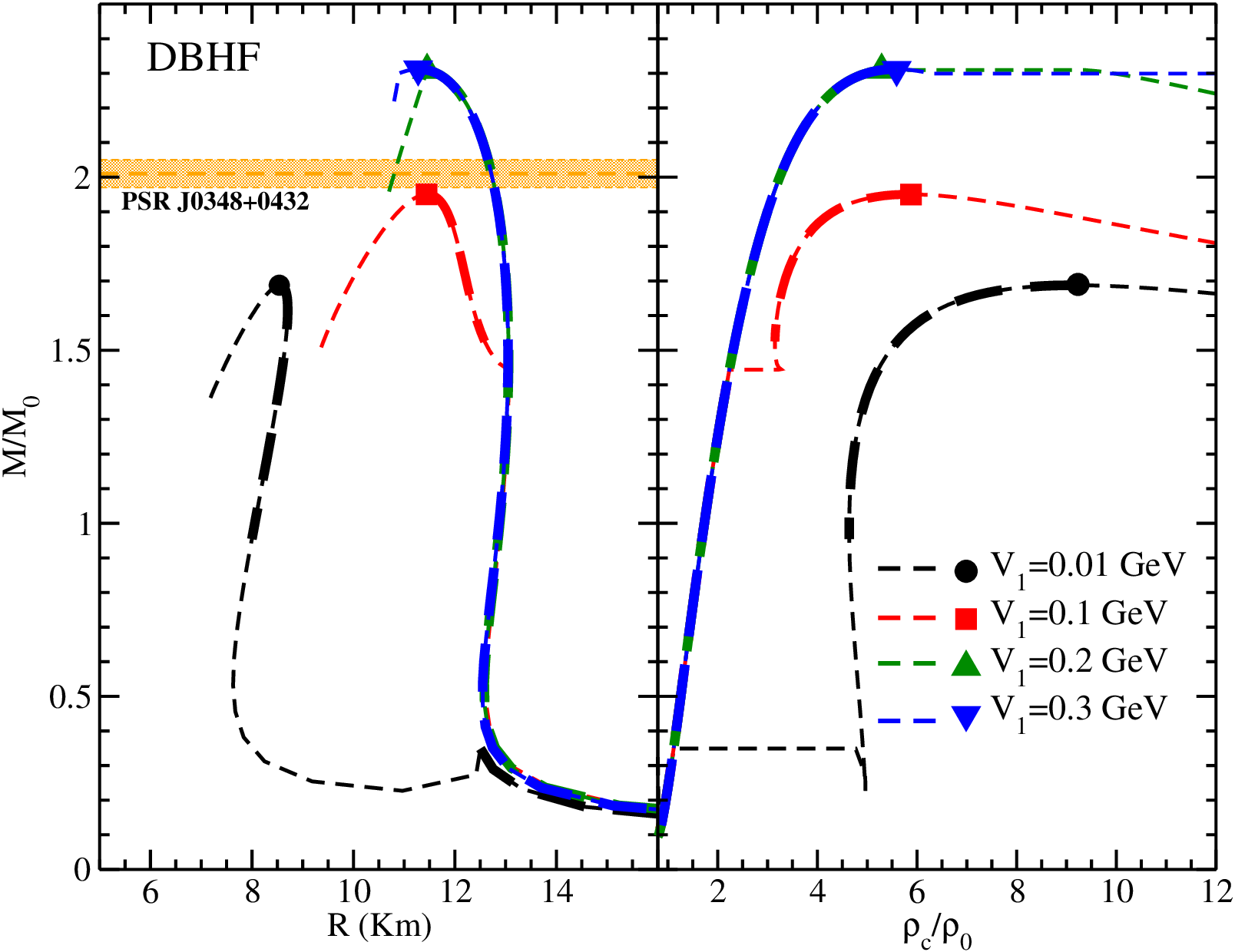}
\caption{Same as Fig.\ref{f:mrbhf_g2} but with DBHF EoS used for the hadronic phase. 
The full symbols denote the value of the maximum mass. (see text for details). }
\label{f:mrdbhf_g2}
\end{figure} 

Similar results are displayed in Fig.\ref{f:mrdbhf_g2}, where the DBHF EoS is used for the 
hadronic phase, with the same notations and coding adopted in Fig.  \ref{f:mrbhf_g2}.
Also in this case  the observational data indicate that values of $\rm V_1$ as small as  
$\rm 0.01 ~GeV$ are excluded, 
and that values of about $\rm 0.1~GeV$ are only marginally compatible with the observational data. 
This is due to the fact that for $\rm V_1 = 0.1~GeV$ the phase transition takes place over a range 
of densities where both BHF and DBHF EoS's show a similar behavior, but in the latter case 
a stable quark matter phase is produced, as can also be deduced from the asterisks reported in Table \ref{t:mass}.

By increasing the value of $V_1$, the value of the maximum mass increases, but  the stability of the pure quark phase is lost, 
as shown in both Figs.\ref{f:mrbhf_g2} and \ref{f:mrdbhf_g2}  
by the cuspids in the mass-radius relation,
and the maximum mass contains in its interior at most a mixed quark-hadron phase.
The maximum mass can increase well above the observational limit with increasing the value
of $\rm V_1$, but hybrid stars will be mainly in the hadronic phase. 
We found that $\rm V_1 \approx 0.12~GeV$ is the largest value which produces stable neutron stars with a quark core if the DBHF EoS is used for the hadronic phase.

\begin{table}[t]
\setlength{\tabcolsep}{5pt}
\caption{Properties of maximum mass configurations. Asterisks denote
stable hybrid stars with pure quark matter core.}
\begin{ruledtabular}
\begin{tabular}{c | c c c c c c c}
&\multicolumn{1}{c}{$\rm V_1 (GeV)$}
& \multicolumn{1}{c}{$\rm M_{max}/M_\odot$}
& \multicolumn{1}{c}{$\rm \rho_c/\rho_0$} 
& \multicolumn{1}{c}{$\rm G_2^c (GeV^4)$} \\
\hline
        & 0.01  &  1.69* & 9.23 & 0.0042 \\
 BHF & 0.1  & 1.95 & 5.47 & $-$ \\
       & 0.2  & 2.03 & 7.24 & $-$ \\
      & 0.3  &  2.03 & 7.24 & $-$ \\
\hline
        & 0.01   &  1.69* & 9.23 & 0.0042 \\
 DBHF  & 0.1   & 1.95* & 5.88 & 0.0023 \\
   & 0.2   &  2.31 & 5.29 & $-$  \\
 	   & 0.3 & 2.31 & 5.59 & $-$ \\
\hline
\end{tabular}
\end{ruledtabular}
\label{t:mass}
\end{table}

\begin{figure}[t] 
\centering
\includegraphics[width=7.5cm,angle=0]{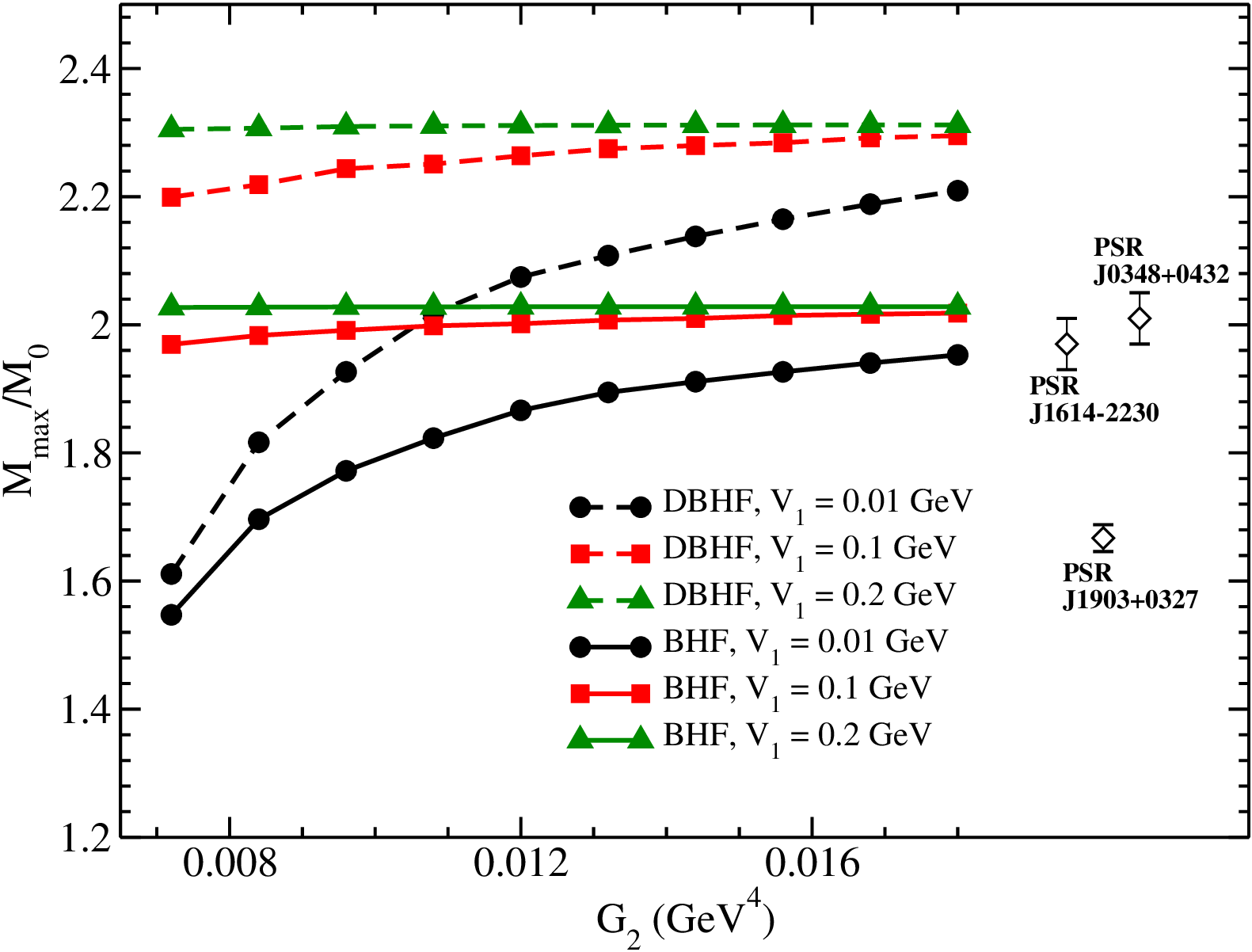}
\caption{The maximum mass, in units of the solar mass $M_\odot$, is displayed vs. the
gluon condensate $G_2$.  $G_2$ is independent on $\mu_B$. }
\label{f:mg2}
\end{figure}

Therefore, generally speaking we can conclude that this model gives values of the maximum
mass in agreement with the current observational data if $\rm V_1 \gtrsim 0.1~ GeV$.
We remind  the reader that those calculations have been performed assuming a dependence of the gluon condensate $G_2$ on the baryon chemical potential  $\mu_B$.  
For comparison, we show in Fig.\ref{f:mg2} the dependence of the maximum mass on the value of  a constant $G_2$, which 
 was discussed in our previous paper \cite{noi08}.  There are some differences,
mainly when a small value of $\rm V_1=0.01~GeV$ is used. In fact, in this case the value of the maximum mass
is compatible with observational data only if a stiff EoS for the hadronic phase is adopted 
and, at the same time,  sufficiently large values of $G_2$ are selected.
With increasing $\rm V_1$, the value of the maximum mass increases, no matter the EoS
for the hadronic phase, as already displayed in Figs. \ref{f:mrbhf_g2} and \ref{f:mrdbhf_g2}with $G_2$ dependent on $\mu_B$. 
Still, the BHF EoS is only marginally compatible with the data whereas the DBHF points lie well above the observed NS masses. 
We also notice that, already at $\rm V_1 = 0.1~GeV$ and more evidently at  
$\rm V_1 = 0.2~GeV$, the values of $\rm M_{max}$ collected in Fig.  \ref{f:mg2} are essentially independent of $G_2$, thus indicating that the 
quark matter appears only after the hadronic branch has reached its maximum,  $\rm M_{max}$,  so that the corresponding star has no quark matter content.

\section{Conclusions}
In this paper we have studied the effects of the appearance of a quark matter core in NS, with the corresponding quark-gluon EoS derived in the framework 
of the FCM and with a suitable parametrization of the gluon condensate in terms of the baryon chemical potential, as
suggested by the analysis of this variable for the theory with  $N_c=2$, where $G_2(\mu_B)$ turns out to be  a decreasing function at small $\mu_B$ and 
increasing at larger $\mu_B$.  The inclusion of a density dependent gluon condensate is  motivated by the expectations of some significant effect 
related to the onset of a superconductive phase at large density, which could be hidden by the use of a constant $G_2$. Therefore, the absence 
of indications coming from lattice simulations at finite $\mu_B$ in the theory with $N_c=3$,  has forced us to resort to the only suggestion available on 
the behaviour of $G_2(\mu_B)$. Clearly these comments put some limits on the quantitative accuracy of the gluon condensate parametrization
here adopted, which however should be regarded just as a  qualitative description that could signal potential flaws of the simpler picture obtained 
by retaining constant $G_2$.

The  same kind of problem shows up for the other parameter entering the quark matter EoS, namely $\rm V_1$, that is expected to decrease 
when $\mu_B$ grows, because of the screening  of the quark-antiquark interaction due to the  increasing density. In this case we 
have no quantitative indication about its $\mu_B$ dependence, except for the suggestion, given in \cite{sim6},  that $\rm V_1$ is 
substantially  $\mu$ independent at least for $\mu\approx 0$. Therefore we performed our analysis at various constant values of 
the potential $\rm V_1$. 

Our results are collected in Figs. \ref{f:mrbhf_g2} and \ref{f:mrdbhf_g2} and Table \ref{t:mass} and they indicate that, 
with the adopted  parametrization of $G_2(\mu_B)$, it is necessary to  have  
$\rm V_1\gtrsim  0.1~GeV$ in order to achieve  $\rm M_{max} \approx 2 M_\odot $.  For larger $\rm V_1$,  $\rm M_{max}$ increases
but no pure quark phase is present in the core of the NS already for $\rm V_1\gtrsim  0.12~GeV$ and, beyond $\rm 0.3~GeV$,  
$\rm M_{max}$ does not grow any more. Its value only depends on the particular hadronic EoS chosen. 
This picture is a substantial refinement of the results found in \cite{noi08} where only vanishing or very small of  $\rm V_1$ were considered.
In fact it was observed that $\rm M_{max}$ grows both with $\rm V_1$ and $G_2$ and in particular it was found $\rm M_{max}=1.78 M_\odot $
at constant $ G_2 = 0.012~{\rm GeV^4}$  and $\rm V_1=  0.01~GeV$, which was consistent with the NS masses measured until then,
but cannot explain the more recent measurements. In addition, in \cite{noi08}, the disappearance of the pure quark phase when increasing $G_2$,
was noticed as well.

Analogous results are also obtained in \cite{bombaci} for  constant $G_2$,  but in this case larger values of  $\rm M_{max}$ are 
obtained because the chosen hadronic EoS, based on the relativistic mean field model, is considerably stiffer than the one adopted in this paper.

A final comment on the  potential $\rm V_1$ is in order. As already noticed, 
our analysis shows that  the observed NS masses require values 
larger than $0.1$ GeV while above $0.3$ GeV 
the quark phase is no longer relevant in the determination of the maximum NS mass.
These characteristic values of $\rm V_1$ are much smaller than those coming from the analysis shown in Sec. \ref{subsec} on 
the lattice data at $\mu_B=0$ around and
above the critical temperature, which, when extrapolated to $T=0$, yield  $\rm V_1(T=\mu_B=0) = 0.8 \div 0.9$ GeV.
This is to be taken as a clear indication that  the long distance potential  is indeed sensitive to the increase of density that induces a screening 
effect.  However we also expect that when $\mu_B$ grows, the quark population increases with respect to the antiquark, so that the quark-quark interaction
becomes more and more relevant, and eventually it is conceivable that it becomes predominant with respect to the quark-antiquark interaction. According to this point one could imagine to replace the singlet, $\rm V_1$, with the antitriplet, $\rm V_3$ and
sextet  $\rm V_6$ interactions, whose relative weights with respect to $\rm V_1$ are respectively $(1/2)$ and $(-1/4)$. 
This leads to an effective interaction whose strength  is about $(1/4)$ of the original one, $\rm V_1(T=\mu_B=0)$, so that 
this extremely simplified picture does nevertheless predict a value of the effective interaction at 
large baryon density ( around $0.2$ GeV), that is 
compatible with the relevant range of $\rm V_1$ suggested by our analysis.

\begin{acknowledgments}
The authors warmly thank M. P. Lombardo (INFN-LNF)  for enlightening discussions concerning
the gluon condensate $G_2$. 
\end{acknowledgments}


\begin{thebibliography}{12}

\bibitem{gabri}
G. Taranto, M. Baldo, and G. F. Burgio,
Phys. Rev. C {\bf 87}, 045803 (2013).

\bibitem{maxpuls}
J. Antoniadis et al., Science {\bf 340}, (2013) 6131.

\bibitem{phrep} A. Di Giacomo, H.G. Dosch, V.I.Shevchenko, and Y.A. Simonov,
Phys. Rep {\bf 372}, (2002) 319.

\bibitem{noi07}
M. Baldo, G.F. Burgio, P. Castorina, S. Plumari, and
D. Zappal\`a, Phys. Rev. C {\bf 75}, (2007) 035804.

\bibitem{noi08}
M. Baldo, G.F. Burgio, P. Castorina, S. Plumari, and
D. Zappal\`a, Phys. Rev. D {\bf 78}, (2008) 063009.

\bibitem{sim1} Yu.A. Simonov, Phys. Lett. B {\bf 619}, (2005) 293.

\bibitem{sim4} Yu.A. Simonov, and M.A. Trusov, JETP Lett. {\bf 85} (2007) 598.

\bibitem{sim22} Yu.A. Simonov, and M.A. Trusov, Phys. Lett. B {\bf 650} (2007) 36.

\bibitem{sim5} Yu.A.Simonov, Annals Phys. {\bf 323} (2008) 783.

\bibitem{sim55} E.V.Komarov, and Yu.A.Simonov, Annals Phys. {\bf 323} (2008) 1230.

\bibitem{sim6}  A. V. Nefediev, Yu.A. Simonov, and M.A. Trusov,  Int. J. Mod. Phys. E {\bf 18} (2009) 549.

\bibitem{latmuf} M. Doring, S. Ejiri, O. Kaczmarek, F. Karsch, and E. Laermann,
Eur. Phys. J.  C{\bf 46} (2006) 179.

\bibitem{gluecond} 
M.A. Shifman, A.I. Vainshtein, and V.I. Zakharov, Nucl. Phys. B {\bf 147} (1979) 385;
Nucl. Phys. B {\bf 147} (1979) 448.

\bibitem{wup_buda}
S. Borsanyi et al., JHEP {\bf 1009} (2010) 073.

\bibitem{wup_buda_2}
S. Borsanyi et al., JHEP {\bf 1011} (2010) 077.

\bibitem{hotqcd}
A.~Bazavov et al.,  Phys. Rev. D {\bf 80} (2009) 014504.

\bibitem{hotqcd_2}
A. Bazavov et al., Phys. Rev. D {\bf 85}, (2012)  054503.

\bibitem{hotqcd_new} 
 USQCD Collaboration, C. De Tar and F. Karsch, "Computational Challenges in QCD Thermodynamics",
  http://www.usqcd.org/documents/13thermo.pdf

\bibitem{elia1} M. D'Elia, A. Di Giacomo, and E. Meggiolaro, Phys. Lett.  B {\bf 408}  (1997) 315.

\bibitem{elia2} M. D'Elia, A. Di Giacomo, and E. Meggiolaro, Phys. Rev. D {\bf 67} (2003) 114504. 

\bibitem{bombaci} I. Bombaci, and D. Logoteta, 
MNRAS {\bf 433} (2013) L79.

\bibitem{cohen} T. D. Cohen, R. J. Furnstahl, and D. K. Griegel, Phys. Rev.  C{\bf 45} (1992) 1881.

\bibitem{druc} E. G. Drukarev, M. G. Ryskin, and V. A. Sadovnikova, Prog. Part. Nucl. Phys. {\bf 47}  (2001) 73.

\bibitem{balcasza} M.Baldo, P.Castorina, and D.Zappal\`a,  Nucl. Phys.  A {\bf  743}, (2004) 3.

\bibitem{zit1} M. A. Metlitski, and A. R. Zhitnitsky, Nucl. Phys. B {\bf  731} (2005) 309.

\bibitem{zit2}  A. R. Zhitnitsky, AIP Conf.Proc.  {\bf 892} (2007) 518.  ArXiv:hep-ph/0701065.

\bibitem{sands} S. Hands, S. Kim, and J. I. Skullerud,  Eur. Phys. J.  C{ \bf 48}  (2006) 193.

\bibitem{alles} B. Alles, M. D'Elia, and M. P. Lombardo, Nucl. Phys. B {\bf 752} (2006) 124.

\bibitem{book}
 M. Baldo,
 {\em Nuclear Methods and the Nuclear Equation of State},
 International Review of Nuclear Physics, Vol. 8
 (World Scientific, Singapore, 1999).

\bibitem{bbb}
 M. Baldo, I. Bombaci, and G. F. Burgio,
 Astron. Astrophys. {\bf 328}, (1997) 274;
 X. R. Zhou, G. F. Burgio, U. Lombardo, H.-J. Schulze, and W. Zuo,
 Phys. Rev. C{\bf 69}, (2004) 018801.
 
 

\bibitem{v18}
 R. B. Wiringa, V. G. J. Stoks, and R. Schiavilla,
 Phys. Rev. C {\bf 51}, 38 (1995).

\bibitem{uix}
 J. Carlson, V. R. Pandharipande, and R. B. Wiringa,
 Nucl. Phys. A {\bf 401}, 59 (1983);
 R. Schiavilla, V. R. Pandharipande, and R. B. Wiringa,
 Nucl. Phys.  A {\bf 449}, 219 (1986).

\bibitem{myers}
 W. D. Myers and W. J. Swiatecki,
 Nucl. Phys. A {\bf 601}, 141 (1996);
 Phys. Rev. C {\bf 57}, 3020 (1998).

\bibitem{fuchs} 
T. Gross-Boelting, C. Fuchs, and A. Faessler,
Nucl. Phys. A {\bf 648}, 105 (1999).

\bibitem{Z_diag} G. E. Brown, W. Weise, G. Baym, and J. Speth, 
Comments Nucl. Part. Phys. {\bf 17}, 39 (1987).


\bibitem{hypmat}
 H.-J. Schulze, A. Lejeune, J. Cugnon, M. Baldo, and U. Lombardo,
 Phys. Lett. B {\bf 355}, (1995) 21;
 H.-J. Schulze, M. Baldo, U. Lombardo, J. Cugnon, and A. Lejeune,
 Phys. Rev. C {\bf 57}, (1998) 704.

\bibitem{hypns}
 M. Baldo, G. F. Burgio, and H.-J. Schulze,
 Phys. Rev. C {\bf 58},  (1998) 3688;
 Phys. Rev.  C{\bf 61}, (2000) 055801.


\bibitem{shapiro} S.L. Shapiro and S.A. Teukolsky,
 {\it Black Holes, White Dwarfs and Neutron Stars}
(John Wiley and Sons, New York, 1983).

\bibitem{fodor} Z. Fodor and S. D. Katz, JHEP {\bf 0404} (2004) 050.

\bibitem{gle}
 N. K. Glendenning, {\em Compact Stars, Nuclear Physics, Particle Physics, and General
Relativity}, 2nd ed. (Springer, New York, 2000).


\bibitem{mixed}
T. Endo, T. Maruyama, S. Chiba, and T. Tatsumi,
Prog. Theor. Phys. {\bf 115}, (2006) 337.

\bibitem{mit}
 G. F. Burgio, M. Baldo, P. K. Sahu, A. B. Santra, and H.-J. Schulze,
 Phys. Lett. B {\bf 526}, (2002) 19;
 G. F. Burgio, M. Baldo, P. K. Sahu, and H.-J. Schulze,
 Phys. Rev. C {\bf 66}, (2002) 025802.

\bibitem{nv}J. W. Negele, and D. Vautherin, Nucl. Phys. A{\bf  207},
(1973) 298.

\bibitem{bps}
 G. Baym, C. Pethick, and D. Sutherland,
 Astrophys. J. {\bf 170}, (1971) 299.
 
 \bibitem{fmt}
 R. Feynman, F. Metropolis, and E. Teller,
 Phys. Rev. C{\bf 75},  (1949) 1561.


\end{thebibliography}
\end{document}